\documentclass{ws-procs9x6}
\usepackage{subfigure}
\begin{document}

\title{Measurements of $C_z$, $C_x$ for $K^+ \Lambda$ and $K^+ \Sigma ^{\circ}$ Photoproduction}

\author{Robert Bradford}

\address{Department of Physics and Astronomy\\
University of Rochester \\
500 Wilson Boulevard\\
Rochester, NY 14627-0171, USA\\ 
E-mail: bradford@pas.rochester.edu}

\author{Reinhard Schumacher}

\address{Dept of Physics\\
Carnegie Mellon University \\ 
5000 Forbes Avenue\\
Pittsburgh, PA 15213\\
E-mail: schumacher@cmu.edu}  

\author{for the CLAS Collaboration.}

\maketitle

\abstracts{The CLAS collaboration has recently completed first measurements of the double polarization observables $C_x$ and $C_z$ for the reactions $\gamma p \rightarrow K^+ \Lambda$ and $\gamma p \rightarrow K^+ \Sigma ^{\circ}$.  $C_x$ and $C_z$ are the beam-recoil polarization asymmetries measuring the polarization transfer from incoming circularly polarized photons to outgoing hyperons along two directions in the reaction's production plane.  The $\Lambda$ is found to nearly maximally polarized along the direction of incident photon's polarization for forward-going kaons.  Polarization transfer to the $\Sigma ^{\circ}$ is different from the $\Lambda$ case.}

\section{Introduction}
Measurement of polarization observables have long been recognized as key to unraveling baryon production mechanisms.  This work represents the first measurement of $C_x$ and $C_z$ for $\gamma p \rightarrow K^+ \Lambda$ and $\gamma p \rightarrow K^+ \Sigma ^{\circ}$.

\section{$C_x$ and $C_z$}
$C_x$ and $C_z$ measure the polarization transfer from a circularly polarized incident photon beam to the recoiling $\Lambda$ or $\Sigma^{\circ}$ baryons along two orthogonal directions in the production plane of the $K^+$-hyperon system.  The $\hat{x}$ and $\hat{z}$ directions are defined in the CM frame, with $\hat{z}$ lying along the directions of the incident photon beam's polarization.  In this analysis, the polarization transfer was measured through the beam helicity asymmetry according to 
\begin{equation}
A_{x/z}\left( \cos \theta _p \right) = \frac{N^+\left( \cos \theta _p \right) - N^-\left( \cos \theta _p \right)}{N^+\left( \cos \theta _p \right) + N^-\left( \cos \theta _p \right)} = \alpha _{eff} \eta C_{x/z} \cos \theta _p
\end{equation}
where $\cos \theta _p$ is the direction of the proton from the hyperon's decay measured in the hyperon rest frame with respect to the $\hat{x}$ or $\hat{z}$ axis.  $\eta$ is the polarization of the incident photon beam, $N^+ \left( \cos \theta _p \right)$ and $N^- \left( \cos \theta _p \right)$ are the beam helicity dependent hyperon yields in a given $\cos \theta _p$ bin.  $\alpha _{eff}$ is the effective weak decay asymmetry parameter, and has a value of 0.642 for $K^+ \Lambda$ and -0.165 for $K^+ \Sigma ^{\circ}$.  The value of $\alpha _{eff}$ for the $\Sigma^{\circ}$ decay arises from our technique of measuring the proton distribution in the rest frame of the $\Sigma^{\circ}$, not the $\Lambda$;  this dilutes its value to less than the nominal -0.642/3.

\section{Experimental Setup and Analysis}
The data were taken using the CLAS spectrometer in Hall B at Jefferson Lab.  The experiment used a circularly polarized photon beam incident on a liquid hydrogen target.  Data were taken with endpoint photon energies of 2.4 and 2.9 GeV.  From this dataset, we also measured differential cross sections, which are currently available in preprint\cite{cspaper}.

All analyzed events were required to have explicit detection of the $K^+$ and proton.  The $\Lambda$ or $\Sigma ^{\circ}$ hyperons were identified in the $p \left( \gamma, K^+ \right)Y$ missing mass.  The data were binned in beam helicity, the cosine of the kaon angle in the CM frame ($\cos \left( \theta _{KCM} \right)$), the cosine of the proton angle ($\cos \left( \theta _p \right)$) and photon energy ($E _{\gamma}$).  Within kinematic each bin, yields were extracted by fitting a Gaussian peak to each hyperon in the missing mass spectrum.  Backgrounds were modeled with a polynomial.  The beam helicity asymmetry was plotted against $\cos \left( \theta _p \right)$ and the slope of this distribution was extracted with a linear fit.  Complete analysis details are available elsewhere\cite{thesis}.

\section{Results and Discussion}
\begin{figure}
\begin{center}
\subfigure[]{
\includegraphics[width=4.0in, height=2.60in]{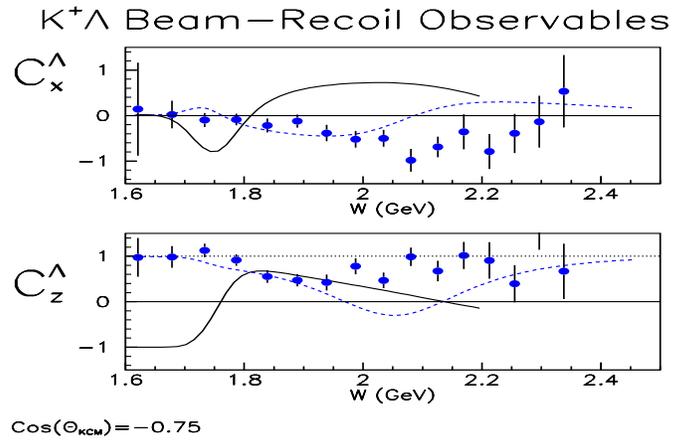}
\label{lam1}
}
\subfigure[]{
\includegraphics[width=4.0in, height=2.60in]{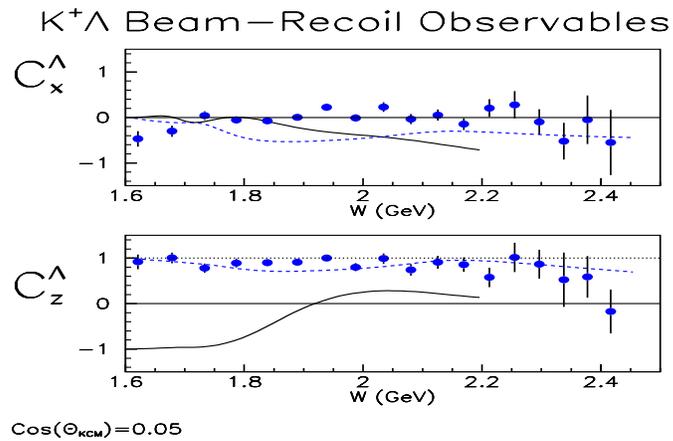}
\label{lam5}
}
\caption[$C_x$ and $C_z$ for $K^+ \Lambda$]{$C_x$ and $C_z$ for $K^+ \Lambda$ in two different kaon-angle bins.  Top:  $\cos \left( \theta _K \right) = -0.75$, bottom: $\cos \left( \theta _K \right) = 0.25$.  The data are a subset of the 2005 CLAS results.  The curves are predictions from the Kaon-MAID (solid line, \cite{maid}) and Janssen (dashed line, \cite{janssen}) isobar models.}
\label{lambda}
\end{center}
\end{figure}

\begin{figure}
\begin{center}
\includegraphics[width=4.0in, height=2.60in]{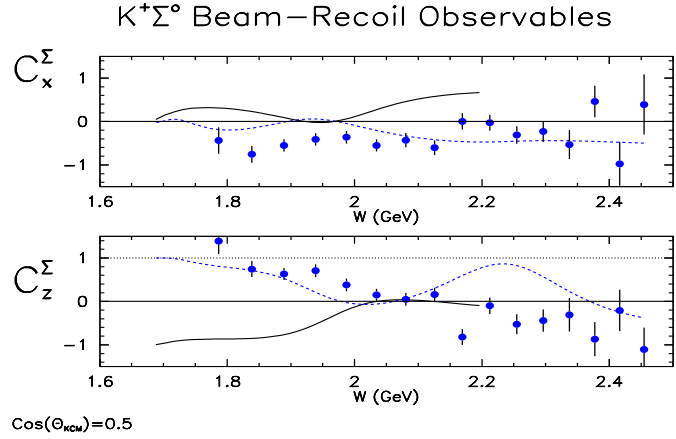}
\caption[$C_x$ and $C_z$ for $K^+ \Sigma ^{\circ}$]{$C_x$ and $C_z$ for $K^+ \Sigma ^{\circ}$ for $\cos \left( \theta _K \right) = 0.5$.  The data are a subset the 2005 CLAS results.  The curves are predictions from the Kaon-MAID (solid line, \cite{maid}) and Janssen (dashed line, \cite{janssen}) isobar models.}
\label{sigma}
\end{center}
\end{figure}

Sample results are presented in Figures \ref{lambda} and \ref{sigma}.  The data are plotted with predictions from the Kaon-MAID\cite{maid} (solid line) and Janssen\cite{janssen} (dashed line) isobar models.  The data plotted are for only a few representative bins in $\cos \left( \theta _{KCM} \right)$.  The full results include nine bins in $\cos \left( \theta _{KCM} \right)$ for $K^+ \Lambda$ and six bins for $K^+ \Sigma ^{\circ}$.
 
The $\Lambda$ results show some W-dependent structure at backward kaon angles and then stabilize at more forward-going kaon angles.  For this hyperon, $C _z$ is near one over most of the forward hemisphere of the kaon angle while $C_z$ is near zero for the same range.

The $K^+ \Sigma^{\circ}$ results show no preferred direction for the polarization over the kaon angle range.  The precision of these results here appears worse due to the small value of $\alpha_{eff}$.
 
Of the models shown, the Janssen does a good job of following the data.  The MAID curve does not fair as well.  This model has the oddity of predicting that $C_z$ saturates at -1 in $K^+ \Lambda$ for forward-going kaons.

\end{document}